\documentclass[twocolumn,showpacs,preprintnumbers,amsmath,amssymb]{revtex4}

\usepackage[dvips]{graphicx}% Include figure files
\usepackage{dcolumn}% Align table columns on decimal point
\usepackage{bm}% bold math

\begin{document}

\title{Rabi-like oscillations of an anharmonic oscillator: classical versus quantum interpretation}

\author{J. Claudon$^{1,2}$, A. Zazunov$^3$, F. W. J. Hekking$^3$, and O. Buisson$^1$}

\affiliation{$^1$Institut N\'eel, C.N.R.S.- Universit\'e Joseph Fourier, BP 166, 38042 Grenoble-cedex 9, France}

\affiliation{$^2$CEA-CNRS joint group 'Nanophysique et Semiconducteurs', CEA, Inac, SP2M, F-38054 Grenoble, France}

\affiliation{$^3$LPMMC, C.N.R.S.- Universit\'e Joseph Fourier, BP 166, 38042 Grenoble-cedex 9, France}

\date{\today}

\begin{abstract}
We have observed Rabi-like oscillations in a current-biased dc SQUID presenting enhanced coherence times compared to our previous realization~\cite{Claudon_PRL04}. This Josephson device behaves as an anharmonic oscillator which can be driven into a coherent superposition of quantum states by resonant microwave flux pulses. Increasing the microwave amplitude, we study the evolution of the Rabi frequency from the 2-level regime to the regime of multilevel dynamics. When up to $3$ levels are involved, the Rabi frequency is a clear signature of quantum behavior. At higher excitation amplitude, classical and quantum predictions for the Rabi frequency converge. This result is discussed in the light of a calculation of the Wigner function. In particular, our analysis shows that pronounced quantum interferences always appear in the course of the Rabi-like oscillations.
\end{abstract}

\pacs{...}
\maketitle

\section{Introduction}

It is well-known that the classical dynamics of an anharmonic oscillator driven by an external monochromatic force can exhibit very complex and rich behavior, such as nonlinear resonances, bistability oscillations or chaotic dynamics\cite{Landau,Linsay_PRL81,Likharev&Barone,Siddiqi_PRL04}. The search for a quantum signature in the quantum counterpart of this problem has triggered a lot of interest~\cite{Milburn_PRL86,Dykman_ZETF88,Quantum_Chaos,Enzer_PRL97,Katz_PRL07,Serban_PRL07}. Ideally, these investigations require a well defined system that can be tuned all the way from the classical to the quantum limit. 

In this perspective, Josephson phase circuits are appealing experimental systems. Their phase dynamics mimics the one of a fictitious particle oscillating in a washboard-like potential well, with associated discrete energy levels. The system anharmonicity is tuned through external bias parameters, such as current or magnetic flux. In a two-level system with a sufficiently long coherence time, Rabi oscillations appear during the transient following the switching on of the microwave driving field \cite{Martinis_PRL02}. A multilevel system also supports similar (Rabi-like) oscillations \cite{Claudon_PRL04}. In this regime, the energy granularity is still strong as compared to the mean energy. Thus the multilevel dynamics was first fully described using a quantum approach. Later, Gronbech-Jensen et al. have reproduced these oscillations using a purely classical theory \cite{Gronbech_PRL05}. Moreover, it was shown that multilevel Rabi-like oscillations were less sensitive to thermal spoiling than their 2-level counterpart \cite{Lisenfeld_PRL07}. These results raise the intriguing question addressed in the present paper: are Rabi-like oscillations an unambiguous proof of quantum behaviour? Besides this fundamental motivation, a fine understanding of the multilevel regime is also of interest for the accurate control of phase qubits, and has motivated recent theoretical \cite{Steffen_PRB03,Amin_LTP06} and experimental works \cite{Meier_PRB05,Strauch_I3E_07,Dutta_AXv08}.

Very recently, it was proposed to analyse the resonance lineshape under strong driving to track classical and quantum signatures \cite{Shevchenko_NJP08}. In this paper, we adopt a complementary time domain-approach. We report the measurement of Rabi-like oscillations in a dc SQUID circuit for various $\mu$w amplitude. Improved coherence times with respect to previous experiments~\cite{Claudon_PRL04} enable us to explore the crossover from two-level to multilevel dynamics in detail. When up to $3$ levels are involved in the oscillation, the experimental Rabi frequency versus the excitation amplitude exhibits a clear quantum behavior, which can definitely not be explained by a classical approach. However, when more than $4$ levels participate in the oscillation, the classical and quantum predictions converge. We discuss this issue in the light of Wigner function calculations, showing in particular that when the system energy reaches its maximum value, quantum interference effects are strong, even in the case of a high power excitation.

\section{The dc SQUID and its operation}

\begin{figure}
\resizebox{0.45\textwidth}{!}{\includegraphics{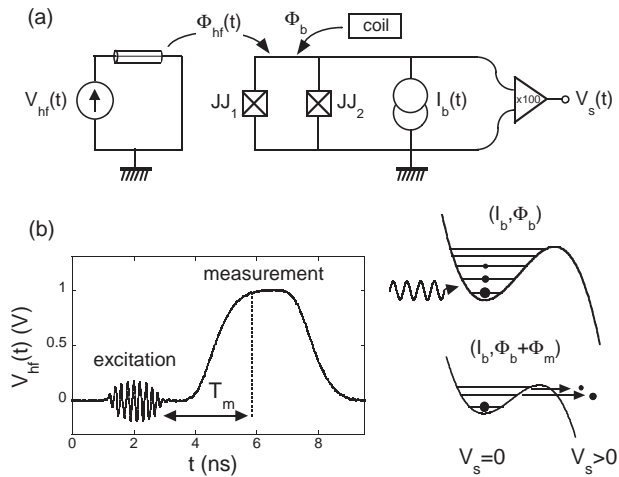}}
\caption{Principle of the SQUID operation. (a) Schematic of the
electrical circuit. (b) Digital sampling oscilloscope record of
the high frequency flux signal applied on the superconducting
loop: the $\mu$w excitation pulse is followed by a measuring dc
pulse which adiabatically reduces the barrier height and induces a
selective tunnel escape of excited states.} \label{circuit}
\end{figure}

The anharmonic oscillator under consideration is realized in a
current biased dc SQUID, which consists of two identical Josephson
junctions (JJ), each with a critical current $I_0$ and a
capacitance $C_0$. The junctions are embedded in a superconducting
loop of total inductance $L_s$, threaded by a flux $\Phi_b$. As
discussed in Ref.~\cite{Balestro_PRL03}, the phase dynamics of the
SQUID can be treated as that of a fictitious particle having a
mass $m=2C_0(\Phi_0/2\pi)^2$ moving in a one-dimensional cubic
potential ($\Phi_0 = h/2e$ is the superconducting flux quantum).
The potential is completely characterized by the frequency $\nu_p$
of the bottom of the well and a barrier height $\Delta U$
[Fig.~\ref{circuit}]. These two quantities depend on the magnetic
flux and vanish at the SQUID's critical current $I_c$. For bias
currents $I_b < I_c$, the particle is trapped in the anharmonic
potential well. In the presence of an applied driving magnetic flux,
the corresponding Hamiltonian reads:
\begin{equation}
\hat{H}_0 (t) =  \frac{\hat{p}^2}{2m}+\frac{1}{2} m\omega_p^2 \hat{x}^2 -a\hat{x}^3 + f_{\mu w} \cos(2\pi \nu t) \hat{x} \,.
\label{Hamiltonian}
\end{equation}
Here, $\omega_p=2\pi \nu_p$; $\hat{x}$ is the phase along the
escape direction, and $\hat{p}$ is the conjugate momentum
associated with the charge on the junction capacitance, $\hat{Q} =
( 2e/\hbar) \hat{p}\,$. The cubic coefficient is given by:
\begin{equation}
a^2=\frac{(m\omega_p^2)^3}{54\Delta U}.
\end{equation}
The $\mu$w flux acts as an
oscillating force with a  strength $f_{\mu w}$ and a frequency
$\nu$. Starting from Eq.~(\ref{Hamiltonian}), the SQUID dynamics
can be treated either classically or with quantum mechanics. In a
quantum description and in the absence of driving flux signal, the
quantized vibration states within the potential well are denoted
$\left| n \right>$, corresponding to energy levels $E_n$ with
$n=0,1,...$ The frequency associated to the $\left| n \right>
\rightarrow \left| k \right>$ transition is denoted $\nu_{nk}$.
The parameter $\delta=\nu_{01}-\nu_{12}$ fully characterizes the system
anharmonicity. At second order in a perturbative approach:
\begin{equation}
\delta=\frac{15}{4 \pi}\frac{\hbar a^2}{m^3 \omega_p^4}=\frac{5}{36} \frac{h \nu_p}{\Delta U} \nu_p,
\label{eq:delta}
\end{equation}
the detuning between the transition $\left| n-1 \right>
\rightarrow \left| n \right>$ and the fundamental one $\left| 0 \right>
\rightarrow \left| 1 \right>$ scales as $\nu_{01}-\nu_{n-1,n}=\delta \times (n-1)$. In the following, for improved precision in the data analysis, the eigenenergies $E_n$ of the system were determined numerically.

Our procedure to perform experiments consists in the repetition of
an elementary sequence, which is decomposed into four successive
steps. First, a bias current $I_b$ is adiabatically switched on
through the SQUID at fixed magnetic flux $\Phi_b$. The working
point $(I_b,\Phi_b)$ defines the geometry of the potential well
and the circuit is initially in the ground state. A fast composite
flux signal, presented in Fig.~\ref{circuit}(b), is then applied.
The $\mu$w excitation  pulse is followed by a dc flux pulse which
brings the system to the measuring point within a few nanoseconds.
This flux pulse reduces adiabatically the barrier height $\Delta
U$ and allows the escape of the localised states to finite voltage
states. Adjusting precisely the amplitude and duration of the dc
flux pulse, it is theoretically possible to induce a selective
escape of excited states. Because the SQUID is hysteretic, the
zero and finite voltage states are stable and the result of the
measurement can be read out by monitoring the voltage $V_s$ across
the dc SQUID. Our measurement procedure destroys the
superconducting state and $I_b$ has to be switched off to reset
the circuit.

Though adiabatic, our measurement is one of the fastest
implemented in Josephson qubits. The delay between the end of
$\mu$w and the measurement (top of the dc pulse) is less than
$3\:\text{ns}$. The measurement time, corresponding to the top of
the pulse, is $4\: \text{ns}$. The measurement speed is crucial to
efficiently detect fast relaxing states. This is especially true
for a multilevel system, since the relaxation rate of the excited
state $\left|n\right>$ scales as $n$. Detection in the multilevel
regime has been characterized in Ref.~\cite{Claudon_PRB07} for
settings identical to those  used in this article. For states
described by the set of occupancies $\{p_0, p_1, p_2, p_3\}$, the
escape probability $P_e$ out of the potential well reads:
$P_e=0.03+0.54p_1+0.79p_2+0.90p_3$.

The SQUID studied in this article consists in two large JJs of
$15\: \mu\text{m}^2$ area ($I_0=1.242\: \mu\text{A}$ and
$C_0=0.56\: \text{pF}$) enclosing a $350\: \mu\text{m}^2$-area
superconducting loop. The two SQUID branches of inductances $L_1$
and $L_2$ contribute to the total loop inductance $L_s=280\:
\text{pH}$ with the asymmetry parameter $\eta = (L_1-L_2)/L_s =
0.414$. As in our previous work, these parameters were determined combining two independent sets of measurements: i) macroscopic quantum tunnelling of the ground state and ii) spectroscopic study of the $\left| 0 \right>
\rightarrow \left| 1 \right>$ transition \cite{Claudon_PRB07}.

The chip is cooled down to $30\: \text{mK}$ in a dilution
fridge. The thermal energy is then small compared to the oscillation
energy of the circuit $h \nu_p \sim k_B \times 500\: \text{mK}$.
The high frequency flux signal is guided by $50\: \Omega$ coax
lines and attenuated at low temperature before reaching the SQUID
through a mutual inductance. The nominal room temperature $\mu$w
amplitude is denoted in the following $V_{\mu w}$. For details on
the circuit fabrication and the experimental setup description, we
refer to Ref.~\cite{Claudon_PRB07}.

\begin{figure}[t]
\resizebox{0.45\textwidth}{!}{\includegraphics{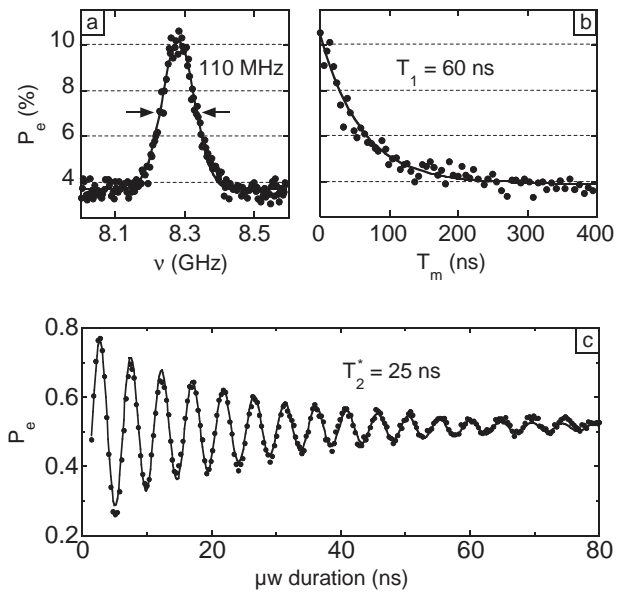}}
\caption{(a): low power spectroscopy ($V_{\mu w}= 17.8\:
\text{mV}$). The fit to a Gaussian lineshape gives the resonance
position and the linewidth. (b): energy relaxation as a function
of the measurement delay $T_m$, fitted to an exponential law with
the damping time $T_1$. (c): Rabi-like oscillation observed in the
transient regime for high power $\mu$w driving ($V_{\mu w}= 631\:
\text{mV}$). The oscillation frequency and the damping time
$T_2^*$ are extracted from a fit to an exponentially damped sine
function.} \label{wp}
\end{figure}

\section{Rabi-like oscillations}

In this section, we present experimental Rabi-like oscillations and their classical and quantum description. Particularly, we focus on the dependence of the Rabi-like frequency on the microwave excitation amplitude: we examine when this particular measurement offers a clear signature of quantum behavior. The observability of beatings, predicted by the quantum approach, is also discussed in the end of the section.

\subsection{Experimental procedure}

Hereafter we present results obtained at the working point $(I_b=2.222\: \mu\text{A}, \Phi_b=-0.117\: \Phi_0$). The
experimental low power spectroscopy measurements (Fig.~\ref{wp}(a)) present a resonant peak centered at $8.283\:
\text{GHz}$ with a full width at half maximum $\Delta \nu = 110\: \text{MHz}$. The resonance is interpreted in a quantum description as the transition from $\left|0\right>$ to $\left|1\right>$ at the frequency $\nu_{01}$. Having the SQUID electrical parameters in hand, the characteristics of the potential well are calculated. The plasma frequency is $\nu_p=8.428\: \text{GHz}$ and the well contains $8$ energy levels for a total depth $\Delta U = h \times 61.6\: \text{GHz}$. Compared to Ref.~\cite{Claudon_PRL04}, the present sample has lower decoherence ($\Delta \nu=110\: \text{MHz}$ instead of $180\: \text{MHz}$), and the chosen bias point displays higher anharmonicity ($\delta=160\: \text{MHz}$ instead of $100\: \text{MHz}$). This improvement allows for much more transition-selective coherent $\mu$w driving.

The transient non linear dynamics of the SQUID is probed using a fast $\mu$w flux pulse followed by a measuring dc pulse. The $\mu$w excitation is tuned on the resonance frequency obtained from low power spectroscopy and the $\mu$w pulse duration is increased from $2\: \text{ns}$ to $80\: \text{ns}$. The measurement delay, as well as
the other measurement pulse settings, remain unchanged. During the transient regime, Rabi-like oscillations (RLO) are observed, as shown in Fig.~\ref{wp}(c). From a fit of the first oscillations to a damped sine function, the RLO frequency $\nu_R$ and the damping characteristic time $T_2^*$ are extracted. It reaches here $25\: \text{ns}$, confirming the coherence improvement in this sample. The dependence of $\nu_R$ with the $\mu$w amplitude is plotted in
Fig.~\ref{fig:rabi}. The error bar on $\nu_R$ is smaller than the experimental point size, around $5\: \text{MHz}$. In our experiments, we have chosen to excite levels which do not leak to voltage states before the measurement pulse. This  implies a limitation on the maximum microwave excitation amplitude. Experimentally, this was checked by performing an experiment without measurement pulse, and by comparing the escape probability with and without microwaves.

\subsection{Classical interpretation}

\begin{figure*}[t]
\resizebox{\textwidth}{!}{\includegraphics{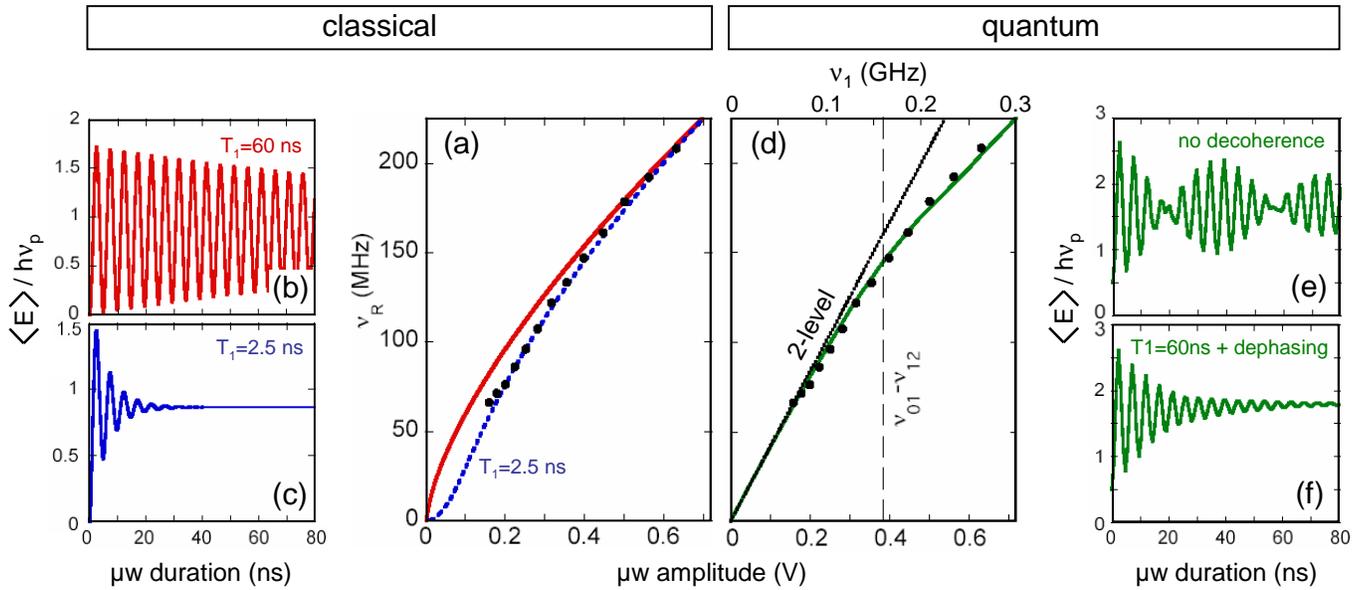}}
\caption{(Color online) Fit of the experimental $\nu_R$ ($\bullet$) to a classical (a) and quantum theory (d). In the case of weakly damped Rabi oscillations, $\nu_R$ is independent of the damping time and identical to the one of a lossless system, both in the quantum and classical cases. The classical model (solid line in (a)) fails to describe the low amplitude regime, unless excessive losses, incompatible with the measured $T_1$ are introduced (dashed line). The quantum prediction (solid line in (c)) accurately describes the experiment over the whole excitation amplitude range. For $\nu_1 > \nu_{01}-\nu_{02}$, the curve deviates from the two level theory (dashed line), indicating a multilevel dynamics. Simulated Rabi oscillation for a $\mu$w amplitude corresponding to Fig.~\ref{wp}(c) are also shown under various hypothesis: classical calculation with $T_1=60\: \text{ns}$ (b) and $T_1=2.5 \: \text{ns}$ (c) and quantum prediction without (e) and with decoherence (f) which suppresses beating.}
\label{fig:rabi}
\end{figure*}

To describe the observed Rabi-like oscillations, we first apply the classical approach developed in Ref.~\cite{Gronbech_PRL05}. Having turned on the resonant driving force, the particle energy oscillates during a transient regime. The underlying mechanism is related to the potential non-linearity, which induces a dephasing between the particle oscillations and the $\mu$w excitation. It makes the $\mu$w alternatively accelerate and brake the particle oscillations within the well. In the general case, $\nu_R$ depends on $f_{\mu w}$, $\nu$ and $T_1$, and the complete set of equations in Ref.~\cite{Gronbech_PRL05} must be solved. Here, the excitation was tuned on the resonance frequency obtained from low power spectroscopy, which corresponds to the bottom well frequency $\nu_p$ in a classical approach. In the limit of small damping ($\nu_R T_1 \gg 1$) which corresponds to the experimental situation, $\nu_R$ is then insensitive to $T_1$ and coincides with the value calculated in the absence of dissipation. In the following analysis, we solved numerically the equation in Ref.~\cite{Gronbech_PRL05} under these two conditions. For the particular situation of small $\mu w$ excitation  ($f_{\mu w}/a \ll 1$), an analytical expression can be derived:
\begin{equation}
\nu_R=\frac{\sqrt{3}}{4} \nu_p \big( 1+2 A^2 \big)^{1/3} \Big( \frac{f_{\mu
w}}{m \omega_p^2 } \Big)^{2/3},
\end{equation}
where $A$ is a dimensionless parameter given by:
\begin{equation}
A=\frac{6a}{m \omega_p^2}=\Big[ \frac{18}{27}\frac{m \omega_p^2}{\Delta U} \Big]^{1/2}.
\end{equation}

Fig.~\ref{fig:rabi}(a) shows the fit of experimental data to the classical theory, the calibration coefficient $c=f_{\mu w}/V_{\mu w}$ being the only free parameter of the model. For consistency, we should note that, in a classical description, the low power resonance presented in Fig.~\ref{wp}(a) is associated with the plasma frequency characterizing the bottom of the well. As spectroscopic data are used to extract the SQUID parameters, the classical analysis leads to slightly different electrical parameters; they are listed in note \cite{note_parametres_C}. In the weakly damped limit, the best adjustment which is obtained for $c=3.50 \times 10^{-22}\: \text{J}.\text{V}^{-1}$ does not describe correctly the data on the full range of excitation amplitudes. At low $\mu$w amplitude, the observed linear dependence is strongly different from the $f_{\mu w}^{2/3}$ law predicted by the classical approach. However, the agreement improves for higher driving amplitude; this convergence will be discussed at the end of the paper.

As a remark, it is possible to fit the data if arbitrary high losses are introduced. Decreasing $T_1$ down to $2.5 \: \text{ns}$ and using $c=2.90 \times 10^{-22}\: \text{J}.\text{V}^{-1}$, one obtains a good agreement between the data and the classical theory (dashed line in Fig.~\ref{fig:rabi}(a)). However, this hypothesis is inconsistent with the measured $T_1=60\: \text{ns}$. It also would lead to overdamped Rabi-like oscillations, with a $T_2^* = 3\: \text{ns}$ (Fig.~\ref{fig:rabi}(c)) which is one order of magnitude smaller than the measured one.

\subsection{Quantum approach}

\subsubsection{Decoherence-free calculation}

The explanation of experimental data requires a full quantum treatment. Since moderate decoherence has a minor impact on the Rabi-like frequency, we first develop a pure Hamiltonian theory \cite{Claudon_PRL04,Strauch_I3E_07,claudon_thesis}.

\begin{figure}[t]
\resizebox{0.45\textwidth}{!}{\includegraphics{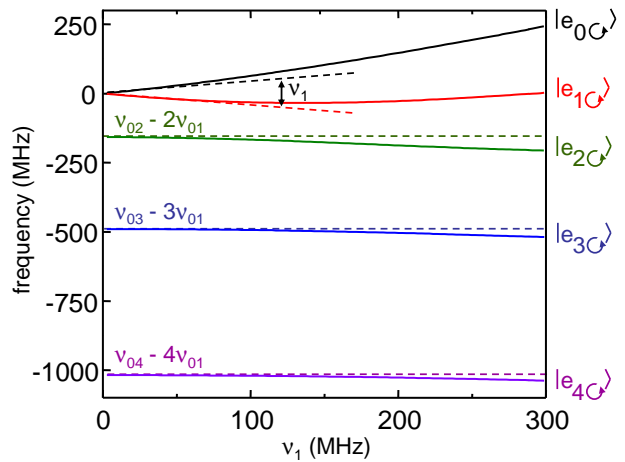}}
\caption{(Color online) First eigenvalues of $\hat{H}_{\circlearrowleft,\text{RWA}}$ as a function of $\nu_1$. The dashed lines correspond to the limit $\nu_1/\delta \ll 1$. The inversion of the spectrum is due to the rotating frame transform.}
\label{fig:propres}
\end{figure}

The particle is assumed to lay initially in the ground state. At $t=0$, the $\mu$w flux is instantaneously switched on, and the particle quantum state $| \Psi (t) \rangle$ evolves into a superposition of the level states $|n\rangle$ according to the Schr\"{o}dinger equation with the Hamiltonian:
\begin{equation}
\begin{aligned}
\hat{H}(t) &=  h \nu_p \Bigg[ \frac{\tilde{p}^2 + \tilde{x}^2}{2} - \Big( \frac{2}{15} \frac{\delta}{\nu_p} \Big)^{\frac{1}{2}} \tilde{x}^3 \Bigg]\\
&+ \sqrt{2} h \nu_1 \cos(2 \pi \nu t)\tilde{x},
\end{aligned}
\label{eq:quantique}
\end{equation}
where $\tilde{x}$ and $\tilde{p}$ are the reduced phase and momentum operators defined by:
\begin{equation}
\tilde{x} = \sqrt{\frac{m \omega_p}{\hbar}} \hat{x} \quad \text{and} \quad \tilde{p} = \frac{1}{\sqrt{m \hbar \omega_p}} \hat{p}.
\end{equation}
The expression Eq.~(\ref{eq:quantique}) contains all the characteristic frequencies of the problem: i) the bottom well frequency $\nu_p$, ii) the detuning $\delta=\nu_{01}-\nu_{12}$ linked to the potential anharmonicity and given by Eq.~(\ref{eq:delta}) and iii) $\nu_1$, the Rabi frequency in the 2-level limit for a microwave excitation frequency $\nu=\nu_{01}$, which reads:
\begin{equation}
\nu_1 =  \nu_p \sqrt{\frac{m \omega_p}{2 \hbar}} \Big( \frac{f_{\mu w}}{m \omega_p^2} \Big).
\end{equation}
To solve this time-dependent problem, we introduce a new picture in which the evolution of the particle state $| \Psi _\circlearrowleft (t) \rangle$ is driven by the Hamiltonian $\hat{H}_\circlearrowleft(t)$. This new picture is related to the Schr\"{o}dinger one by:
\begin{equation}
\langle n | \Psi _\circlearrowleft (t) \rangle = e^{i 2 \pi n \nu t} \langle n | \Psi (t) \rangle.
\label{eq:passage}
\end{equation}
In the 2-level limit, this transformation corresponds to the introduction of the well-known rotating frame at frequency $\nu$. To proceed, we apply the rotating wave approximation (RWA) to simplify the Hamiltonian $\hat{H}_\circlearrowleft(t)$. Considering a potential well with $N$ trapped energy levels, the resulting Hamiltonian $\hat{H}_{\circlearrowleft,\text{RWA}}$ can be expressed in the $\{ |n \rangle \}$ basis as follows:
\begin{equation}
\hat{H}_{\circlearrowleft,\text{RWA}}= h
\left(
\begin{array}{c c c c c}
0                  & \frac{\nu_1}{2}      & \!0                       & \!\!\!0                 \\
\frac{\nu_1}{2} & \Delta_1(\nu)               & \!\ddots                   & \!\!\!0                 \\
0                  & \ddots                  & \!\ddots                  & \!\!\!\sqrt{N\!\!-\!\!1}\frac{\nu_1}{2}   \\
0                  & 0                       & \!\!\sqrt{N\!\!-\!\!1}\frac{\nu_1}{2}     & \!\!\Delta_{N-1}(\nu)  \\
\end{array} \right).
\end{equation}
Here, $\Delta_n(\nu)=\nu_{0n}-n\nu$ and the ground state energy $E_0$ has been set equal to $0$. Note that $\tilde{x}$ presents matrix elements of the order of $\sqrt{\delta/\nu_p}$ on the diagonal and other which connect neighbor states of order $2$. Other terms of higher order in $\sqrt{\delta/\nu_p}$ are also present in the matrix. These terms vanish in the RWA, which remains valid as long as $\nu_1 \ll \nu_{01}$. In this quantum approach, the experimental low power resonance is attributed to the $\left|0\right> \rightarrow \left|1\right>$ transition; thus the following analysis is performed for $\nu=\nu_{01}$. The eigenvalues $\lambda_n$ associated to the eigenstates $| e_{n \circlearrowleft} \rangle$ of $\hat{H}_{\circlearrowleft,\text{RWA}}$ are shown in Fig.~\ref{fig:propres}. Having solved the time evolution of $| \Psi _\circlearrowleft (t) \rangle$ starting from the initial state $| 0_\circlearrowleft \rangle = | 0 \rangle$, the conversion to the Schr\"{o}dinger picture is done according to Eq.~(\ref{eq:passage}). Particularly, the probability to find the system in the state $|n\rangle$ in the course of the oscillation reads for $n>0$:
\begin{equation}
\begin{aligned}
p_n(t) = 2 \sum_{k,l>k}^{N-1} &| \langle n | e_{k \circlearrowleft} \rangle \langle e_{k \circlearrowleft} | 0 \rangle \langle 0 | e_{l \circlearrowleft}  \rangle \langle  e_{l \circlearrowleft} | n \rangle | \\
&\times \cos[2\pi(\lambda_k-\lambda_l)t].
%p_n(t) = \sum_{k,l=0}^{N-1} \langle n | e_{k \circlearrowleft} \rangle \langle e_{k \circlearrowleft} | 0 \rangle \langle 0 | e_{l \circlearrowleft}  \rangle \langle  e_{l \circlearrowleft} | n \rangle e^{-i 2 \pi(\lambda_k-\lambda_l)t}.
\end{aligned}
\label{eq:multi}
\end{equation}
The occupation dynamics of state $| n \rangle$ implies the frequency set $\{ | \lambda_k-\lambda_l | \}$ $(k \ne l)$ with weights associated to the connection of $| 0 \rangle$ to  $| n \rangle$ via the eigenstates of $\hat{H}_{\circlearrowleft,\text{RWA}}$. As shown in Fig.~\ref{fig:rabi}(e), the average energy of the particle $\langle \hat{H} \rangle = \sum _n p_n(t)E_n$ undergoes Rabi-like oscillations. The model also predicts modulations of the RLO amplitude which will be discussed in the next paragraph. Our analysis shows that an estimate of $\nu_R$ can be obtained by taking the minimum value of the frequency set $\{ | \lambda_k-\lambda_l | \}$. For improved precision, the theoretical Rabi-like frequency is extracted from a fit of the first oscillations to an exponentially damped sine function. This approach was checked to give similar results as a direct numerical integration of the time dependent Schr\"{o}dinger equation.

As shown in Fig.~\ref{fig:rabi}(d), the quantum model describes all the experimental features of the $\nu_R$ versus $f_{\mu w}$ dependence with $c=3.05 \times 10^{-22}\: \text{J}.\text{V}^{-1}$. The set of experimental data covers the crossover between the 2-level and the multilevel dynamics. Indeed, when the $\mu$w amplitude is small compared to the anharmonicity ($\nu_1 \ll \delta$), we have $| e_{0 \circlearrowleft} \rangle = (| 0  \rangle + | 1  \rangle)/\sqrt{2}$ , $| e_{1 \circlearrowleft} \rangle = (| 0  \rangle - | 1  \rangle)/\sqrt{2}$ and $| e_{n \circlearrowleft} \rangle = | n  \rangle$ for the other higher energy levels. The dynamics thus concerns the first two levels and one retrieves the familiar result $\nu_R = \nu_1$ [see Fig.~\ref{fig:propres}]. This is nearly the case for the first measured point, which corresponds to $\nu_1 = 65\: \text{MHz}$ and presents a contamination of level $\left| 2 \right>$ below $10\: \%$. When $\nu_1 \sim \delta$, the coupling between neighboring levels starts to distort the energy spectrum of the lowest energy levels of $\hat{H}_{\circlearrowleft,\text{RWA}}$ [Fig.~\ref{fig:propres}]. Consequently, the state $| e_{2 \circlearrowleft} \rangle$ gets contaminated by $| 0  \rangle$ and $| 1  \rangle$ and the dynamics involves now 3 levels. In this regime, the $\nu_R$ dependence on $\nu_1$ starts to deviate from a linear behavior, indicating the onset of two-photon processes \cite{Strauch_I3E_07,Dutta_AXv08}. When $\nu_1 > \delta$, an increasing number of levels are involved. As an example, the experimental oscillation presented in Fig.~\ref{wp}(c) involves $4$ states. At larger $\mu$w amplitude, a macroscopic number of levels are involved; this regime is not achieved at this working point but was discussed in Ref.~\cite{Claudon_PRL04} with about $10$ levels involved. The multilevel dynamics is characterized by a clear saturation of the $\nu_R$ dependence, compared to the low power linear behavior.

\subsubsection{Beating and decoherence}

In the absence of dissipation, the quantum model predicts a modulation of the multilevel RLO amplitude. We interpret this phenomena as a reminiscence of the different frequencies involved in the system dynamics. The modulation becomes more and more complex with increasing $\mu$w amplitude. Indeed, the higher the amplitude, the larger the number of involved levels and the larger number of frequencies in the system. A clear beating appears when the frequency set $\{ | \lambda_k-\lambda_l | \}$ displays two minimal values close to each other.

However, these modulations were not observed in our experiments. We first analyzed the possibility of beating suppression caused by the finite rise time of the $\mu$w amplitude. If it is too slow, the population of the states $| e_{n \circlearrowleft} \rangle$ for $n>2$ can be severely decreased as compared to the instantaneous hypothesis. In the limiting case of an adiabatic increase of the $\mu$w amplitude, the particle state remains in the $\{| e_{0 \circlearrowleft} \rangle, | e_{1 \circlearrowleft} \rangle \}$ subspace. $|\lambda_1-\lambda_0|$ is then the only frequency involved in the system dynamics (see Fig.~\ref{fig:propres}) and there is no beating. For a rise time as large as $2\: \text{ns}$ (twice the experimental rise time), we checked numerically that this effect does not lead to destructive interference strong enough to suppress the beating.

We therefore argue that the strong suppression of the beating phenomenon could be due to decoherence. The dissipative dynamics of the oscillator was investigated using a phenomenological model. We assume that the reduced density matrix of the driven oscillator $\hat{\rho}$ satisfies the master equation
\begin{equation}
i \hbar \partial_t  \hat{\rho} =  [ \hat{H}_{\circlearrowleft,\text{RWA}} , \hat{\rho} ] - i {\cal L}_R[\hat{\rho}] - i {\cal L}_D[\hat{\rho}] \,.
\label{DMeq}
\end{equation}
Here, ${\cal L_R}$ describes relaxation (accompanied by dephasing) and can be expressed as follows:
\begin{equation}
{\cal L_R}[\hat{\rho}]_{kn} =
\sum_m \left[ \left( \gamma_{km} + \gamma_{nm} \right) \rho_{kn} - 2 \delta_{kn} \gamma_{mk} \rho_{mm}\right] \,.
\end{equation}
The coefficients $\gamma_{kn}$ can be calculated microscopically in the Markovian limit, assuming the environment is a bath of harmonic oscillators. The last term ${\cal L_D}$ describes the pure dephasing:
\begin{equation}
{\cal L_D}[\hat{\rho}]_{kn} = (1-\delta_{kn}) \Lambda_{kn} \rho_{kn} \,,
\end{equation}
with $\Lambda_{kn} = \Lambda_{nk} \geq 0$. This term is phenomenological: the coefficients
$\Lambda$ were obtained microscopically only in the two-level
limit. According to our analysis, relaxation alone (corresponding
to $T_1 = 60$ ns in the experiment) is too weak to explain a
strong suppression of the beating. Instead, the absence of beating
can be attributed to strong pure dephasing between
next-neighboring levels, described by $\Lambda_{n,n+2}$. Indeed,
the observed exponential decay of the Rabi-like oscillations with
the suppressed beating can be well explained assuming
$\Lambda_{02} \approx h \alpha/\nu_1$, which becomes relevant for
the measured escape probabilities corresponding to the cases of
$\nu_1 = 130$ and 260 MHz with $\alpha=5\times10^ 6 \: \text{GHz}^2$.

\section{Discussion}

\begin{figure*}[t]
\resizebox{\textwidth}{!}{\includegraphics{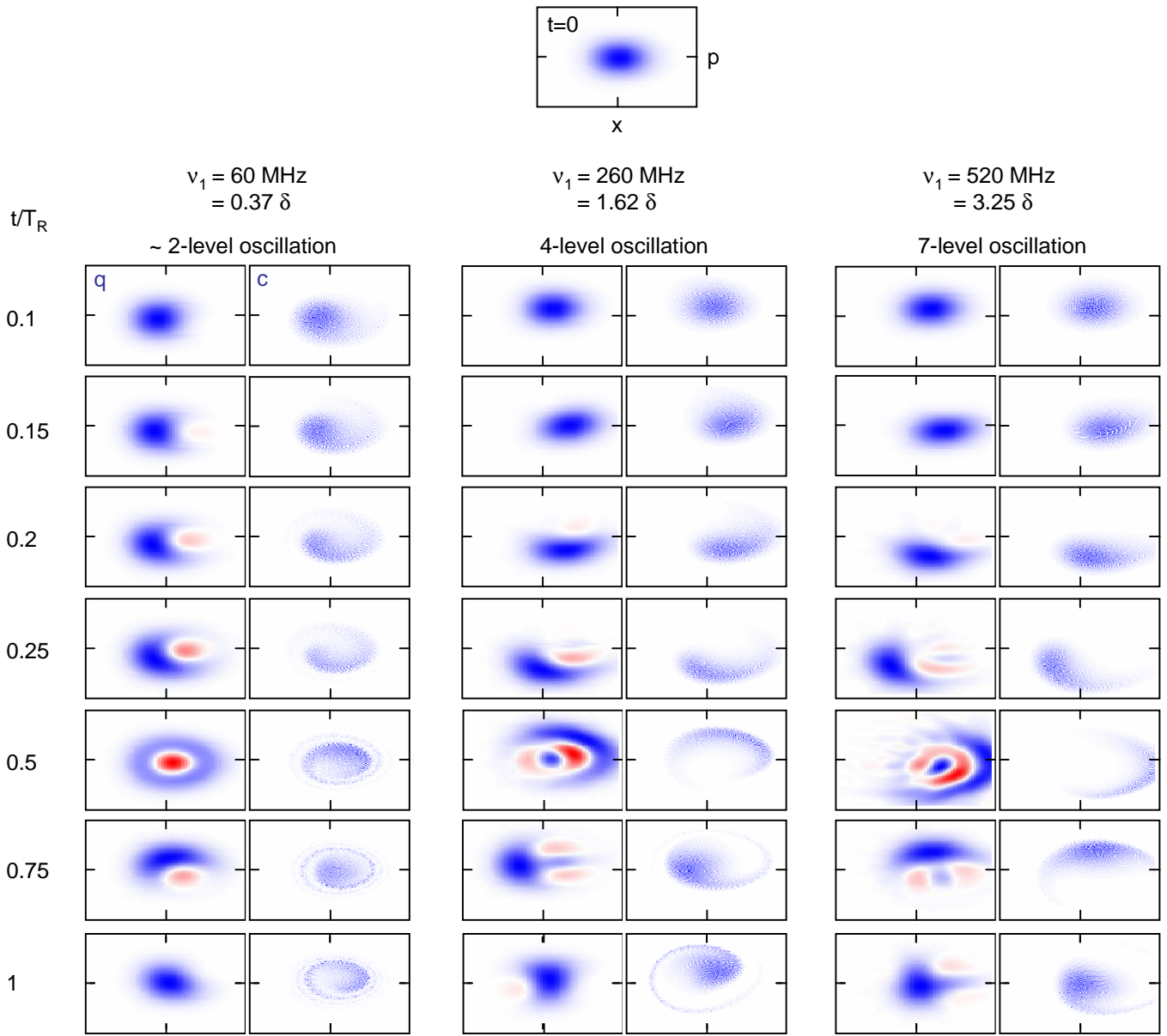}}
\caption{Quantum (q) and classical (c) dynamics during the first Rabi period $T_R$, for increasing excitation amplitudes $\nu_1$. Decoherence is neglected in all calculations. Wigner and classical distributions are plotted in the phase space (position $x$, conjugate moment $p$). In the Wigner distribution plots, red (blue) corresponds to negative (positive) values. The ticks indicate the position of the origin of the phase space. Negative areas are always intense close to $t=0.5 T_R$, when the system experiences fully the non linearity of the potential well, highlighting the quantum nature of the corresponding states. Note also the progressive convergence, as $\nu_1$ increases, between the principal blue cloud of the Wigner map and the classical prediction.}
\label{fig:wigner}
\end{figure*}

Surprisingly, when $\nu_1 > 1.5 \delta$, the classical and quantum predictions for the $\nu_R(\nu_1)$ dependence are quite identical. For the lowest excitation amplitudes meeting this condition, only $4$ levels are involved in the Rabi-like oscillation: the granularity of energy as compared to the mean energy of the system is clearly not negligible. Moreover, this convergence is obtained for weakly damped oscillation, for which moderate relaxation and dephasing do not affect $\nu_R$. Thus, this effect is not due to decoherence which would degrade the system quantum dynamics into a classical one. The convergence is linked to the intrinsic physics of a weakly anharmonic oscillator.

To shed some light on the interplay between non-linearity and the quantum or classical nature of the Rabi-like oscillations, we developed an approach similar to the one presented in Ref.~\cite{Katz_PRL07}. We calculated the time evolution of the Wigner function
\begin{equation}
W(x,p,t) = \frac{1}{\pi \hbar} \int \text{d}x' \langle x+x'|\hat{\rho}(t)|x-x'\rangle e^{-2 i p x'/\hbar},
\end{equation}
during one Rabi oscillation, starting from the oscillator ground state $|0\rangle$. Here, $\hat{\rho}$ is the reduced density matrix describing the driven oscillator. For the sake of comparison, we also determined the classical time evolution of the probability density for a gaussian distribution of initial points in phase space which mimics the distribution associated to the ground state $|0\rangle$. Decoherence effects were neglected in both approaches. This analysis highlights the differences in the system dynamics arising from quantum mechanics. Fig.~\ref{fig:wigner} shows the results for three different excitation amplitudes: i) the lowest excitation amplitude measured in this paper, $\nu_1=60\: $MHz which induces a quasi 2-level oscillation, ii) $\nu_1=260\: $MHz, the highest excitation amplitude and iii) $\nu_1=520\: $MHz, a (non measured) value which generates 7-level oscillation, similar to the high excitation results presented in Ref.~\cite{Claudon_PRL04}.

The red areas in Fig.~\ref{fig:wigner} correspond to negative values of the Wigner function, an unambiguous signature of the quantum nature of the state under consideration; in the blue regions $W$ is positive. As expected for short times (shorter than the so-called Ehrenfest time), the  system behaves classically and the positive part of $W$ resembles the classical probability distribution, regardless of the value of the microwave amplitude, as illustrated in Fig.~\ref{fig:wigner} for $t= 0.1 T_R$. For longer times, this resemblance gets lost for small microwave amplitudes. This is to be expected: due to the anharmonicity, the system remains close to the two-level limit for small amplitudes and the time evolution is essentially quantum. For larger amplitudes, the dynamics involves an increasing number of quantum states and according to the correspondence principle should behave more and more classically. Indeed, the time evolution of the blue regions closely follows that of the classical probability distribution in this limit.

Interestingly, though, is the fact that the red regions of negative $W$ continue to exist at any value of microwave amplitude. These regions, signalling the occurrence of quantum interferences, appear to be most pronounced for $t=0.5 T_R$. A qualitative explanation for this fact can be obtained by considering the time evolution of the anharmonic oscillator, starting from its ground state $|0\rangle$ at time $t=0$. Under the influence of the microwaves, resonant with the transition between the lowest eigenstates, transitions between neighboring levels are possible, and a superposition state is formed involving an increasing number of energy levels. Due to anharmonicity, the detuning increases with increasing level index and the transitions become less resonant. At $t=0.5 T_R$, non-linearity prevents higher energy states from being excited and microwaves lead to a reduction of the system energy. Thus the existence of Rabi-like oscillation is closely linked to the anharmonicity of the potential well and the maximum number of states involved in the superposition is determined by the interplay of the anharmonicity and the microwave-amplitude. Such oscillations do not show up in a harmonic oscillator driven by resonant monochromatic excitation. In this last situation, the generated quasiclassical states present minimum fluctuations and are thus very close to classic ones. On the contrary, in the case of an anharmonic oscillator, for times close to $t=0.5 T_R$, non-linearity gives birth to pronounced quantum interferences, and the corresponding Wigner function is characterized by the existence of large negative regions.

Finally, we can remark that the analysis of the Rabi oscillation frequency is one measurement among many possibilities. It appears that this quantity is not very sensitive to the quantum interferences characterising the extreme state of the Rabi oscillation. To track a quantum signature in the multilevel regime ($\nu_1 > \delta$), one could concentrate on the oscillation beating, a reminiscence of the different transition frequencies in the system. Another appealing perspective would be to extend the state tomography scheme developed in phase Qubits \cite{Steffen_PRL06} to our multilevel system. Though experimentally challenging, this would open interesting possibilities such as the investigation of the effects of decoherence on the state superposition generated at high excitation power.

\section{Conclusion}

In conclusion, a dc SQUID is a tunable and well defined anharmonic quantum oscillator which appears as an experimental model system to explore quantum and classical dynamics. The dependence of the Rabi-like oscillation frequency on the $\mu$w amplitude is a clear probe of the nature of oscillator dynamics when up to $3$ levels participates in the oscillation. It then exhibits a clear quantum behavior which is not described by a classical model. At higher excitation power, when more than $4$ states are involved in the oscillation, the quantum and classical predictions for this particular quantity converge. This convergence was discussed with the help of Wigner quasi-distribution of probability. However, one should not hastily conclude than this non linear oscillator behaves classically as soon as $4$ levels are excited. Indeed, our theoretical study shows that for times corresponding to the middle of the Rabi-like oscillation (the generalization of a $\pi$ pulse), the system always presents strong quantum interferences, even in the high power excitation regime. Future work will concentrate on the shape of the oscillations which contains more information on the multilevel dynamics.

We thank F. Faure, W. Guichard, L. P. L\'evy, and A. Ratchov for
fruitfull discussions. This work was supported by two ACI
programs, by IUF and IPMC and by the EuroSQIP project.

\end{document}